\documentstyle[12pt,axodraw]{article}

\catcode`@=12
\begin{document}
\thispagestyle{empty}
\begin{flushright} UCRHEP-T246\\February 1999\
\end{flushright}
\vspace{1.0in}
\begin{center}
{\Large \bf Mass Splitting of Three Seesaw Neutrinos\\}
\vspace{1.0in}
{\bf Ernest Ma\\}
\vspace{0.3in}
{\sl Physics Department, University of California, 
Riverside,\\ CA 92521, USA\\} 
\vspace{1.0in}
\end{center}
\begin{abstract}\
If a family symmetry exists so that all three neutrinos 
have equal Majorana masses (via the seesaw mechanism), then the breaking of 
this symmetry from charged-lepton masses (via two-loop double $W$ exchange) 
implies $\Delta m$ less than about $10^{-9}~m_0$.  With the common mass 
$m_0 \sim$ eV for hot dark matter, $\Delta m^2 \sim 10^{-10}$ eV$^2$ 
is natural for vacuum solar neutrino oscillations.
\end{abstract}
\vspace{0.5in}
\noindent -----------------

\noindent Talk presented at the 17th International Workshop on Weak 
Interactions and Neutrinos (Cape Town), 1999.

\newpage
\baselineskip 24pt

\section{Introduction}

The excitement generated by the Super-Kamiokande evidence of atmospheric 
neutrino oscillations \cite{1} has prompted a great deal of theoretical 
activity in building models which explain it, as well as the possible 
evidence of solar neutrino oscillations \cite{2}.  With three known 
neutrinos, it is difficult to accommodate the LSND (Liquid Scintillator 
Neutrino Detector) results \cite{3} as well, without sacrificing some 
pieces of the experimental data.  Reviews of the overall situation 
are given in various other talks of these Proceedings.  I will focus 
on the possibility of three nearly mass-degenerate neutrinos.

\section{Common Mass for Three Neutrinos}

Since neutrino oscillations only probe the difference of mass-squares, it 
is entirely possible that the three known neutrinos have a common large 
mass, but their splittings are small for some reason.  This idea has 
received a lot of attention in the recent literature \cite{4} and is being 
pursued actively at present \cite{5}. 

A particularly desirable $(\nu_e, \nu_\mu, \nu_\tau)$ mass matrix for 
oscillations, dark matter \cite{6}, and the absence of neutrinoless double 
beta decay \cite{7} is given by
\begin{equation}
{\cal M} \simeq \left[ \begin{array} {c@{\quad}c@{\quad}c} 0 & -1/\sqrt 2 & 
-1/\sqrt 2 \\ -1/\sqrt 2 & 1/2 & -1/2 \\ -1/\sqrt 2 & -1/2 & 1/2 \end{array} 
\right] ~m_0,
\end{equation}
where the zero $\nu_e$ diagonal entry ensures the absence of neutrinoless 
double beta decay.  The eigenvalues of $\cal M$ are $-m_0$, $m_0$, and $m_0$, 
and the mass eigenstates are related to the interaction eigenstates by
\begin{equation}
\left[ \begin{array} {c} \nu_e \\ \nu_\mu \\ \nu_\tau \end{array} \right] 
\simeq \left[ \begin{array} {c@{\quad}c@{\quad}c} 1/\sqrt 2 & 1/\sqrt 2 & 0 
\\ 1/2 & -1/2 & -1/\sqrt 2 \\ 1/2 & -1/2 & 1/\sqrt 2 \end{array} \right] 
\left[ \begin{array} {c} \nu_1 \\ \nu_2 \\ \nu_3 \end{array} \right].
\end{equation}

Most model builders now add splitting so that
\begin{equation}
(\Delta m^2)_{13} \simeq (\Delta m^2)_{23} \sim 10^{-3}~{\rm eV}^2
\end{equation}
to obtain $\nu_\mu \to \nu_\tau$ oscillations with $\sin^2 2 \theta = 1$. 
Further splitting is then added so that
\begin{equation}
(\Delta m^2)_{12} \sim 10^{-10}~{\rm eV}^2
\end{equation}
to obtain $\nu_e \to (\nu_\mu + \nu_\tau)/\sqrt 2$ oscillations with 
$\sin^2 2 \theta = 1$ also.  Note that $m_0 \sim$ eV rules out the possibility 
of the small-angle matter-enhanced solution of the solar neutrino deficit 
in this context.

\section{Natural Splitting of Three Seesaw Neutrinos}

Whatever mechanism is assumed to obtain three mass-degenerate neutrinos, 
it is broken explicitly by charged-lepton masses.  Hence there must be 
splitting among them from the physics of the standard model, supplemented 
only by the above-mentioned mass-generating mechanism.  Consider thus the 
simplest such extension of the standard model.  Add three heavy right-handed 
neutrino singlets $N_{iR}$ as usual.  Assume that they and the left-handed 
lepton doublets $(\nu_i,l_i)_L$, with $i = +,0,-$, form $SO(3)$ triplets. 
This yields the following invariant terms:
\begin{equation}
f [(\bar \nu_+ N_+ + \bar \nu_0 N_0 + \bar \nu_- N_-) \bar \phi^0 - (\bar 
l_+ N_+ + \bar l_0 N_0 + \bar l_- N_-) \phi^-],
\end{equation}
and
\begin{equation}
M (2 N_+ N_- - N_0 N_0).
\end{equation}
In the basis $(\nu_+, \nu_-, \nu_0, N_+, N_-, N_0)$, the mass matrix is then
\begin{equation}
{\cal M}_{\nu,N} = \left[ \begin{array} 
{c@{\quad}c@{\quad}c@{\quad}c@{\quad}c@{\quad}c} 0 & 0 & 0 & m_D & 0 & 0 
\\ 0 & 0 & 0 & 0 & m_D & 0 \\ 0 & 0 & 0 & 0 & 0 & m_D \\ m_D & 0 & 0 & 0 & M 
& 0 \\ 0 & m_D & 0 & M & 0 & 0 \\ 0 & 0 & m_D & 0 & 0 & -M \end{array} 
\right],
\end{equation}
where $m_D = f \langle \phi^0 \rangle$.  Let $m_D << M$, the well-known 
seesaw mechanism \cite{8} then reduces the above to a $3 \times 3$ mass matrix 
for $(\nu_+, \nu_-, \nu_0)$:
\begin{equation}
{\cal M}_\nu = \left[ \begin{array} {c@{\quad}c@{\quad}c} 0 & -m_0 & 0 \\ 
-m_0 & 0 & 0 \\ 0 & 0 & m_0 \end{array} \right],
\end{equation}
where $m_0 = m_D^2/M$.

Choose $l_+ = e$, then in general $l_- = c \mu + s \tau$ 
and $l_0 = c \tau - s \mu$, where of course $c = \cos \theta$ and $s = \sin 
\theta$.  The degeneracy of ${\cal M}_\nu$ is now lifted through the two-loop 
diagram \cite{9} of Fig.~1.  The result is \cite{10}
\begin{equation}
{\cal M}_\nu = \left[ \begin{array} {c@{\quad}c@{\quad}c} 0 & -m_0-s^2I & 
-scI \\ -m_0-s^2I & 0 & scI \\ -scI & scI & m_0 + 2c^2I \end{array} \right],
\end{equation}
where
\begin{equation}
I = {g^4 \over 256 \pi^4} {m_\tau^2 \over m_W^2} \left( {\pi^2 \over 6} - 
{1 \over 2} \right) m_0 = 3.6 \times 10^{-9} m_0.
\end{equation}
The new eigenvalues are $-m_0-s^2I$, $m_0$, and $m_0 + (1+c^2)I$.  For small 
$s^2$, the above yields $\nu_e \to \nu_\mu$ oscillations with $\sin^2 2 \theta 
\simeq 1$ and $\Delta m^2 = 2 s^2 m_0 I$ which is about $3 \times 10^{-10}$ 
eV$^2$ for $s = 0.1$ and $m_0 = 2$ eV.  This shows that vacuum solar neutrino 
oscillations are natural with three seesaw neutrinos of the same tree-level 
mass.

\section{Atmospheric Neutrino Oscillations}

It has been shown in the above that with equal tree-level seesaw masses for 
the three Majorana neutrinos, there is an irreducible splitting among them 
which is of the right magnitude \cite{11} for vacuum solar neutrino 
oscillations.  However, there is no explanation of atmospheric neutrino 
oscillations in this minimal version of the model.  An {\it ad hoc} 
assumption may be made that the state $c' \nu_0 + s' (\nu_+ - \nu_-)/\sqrt 2$ 
acquires a mass $m_1$ with $m_1/m_0 \simeq 5 \times 10^{-4}$.  In that case, 
with $s' << 1$ but not zero, the mass eigenvalues become $-m_0-s^2I$, 
$m_0 + (s^2 + 2 \sqrt 2 s' sc)I$, and $m_0 + m_1$, with
\begin{equation}
\left[ \begin{array} {c} \nu_e \\ \nu_\mu \\ \nu_\tau \end{array} \right] 
\simeq \left[ \begin{array} {c@{\quad}c@{\quad}c} 1/\sqrt 2 & 1/\sqrt 2 & 
s'/\sqrt 2 \\ c/\sqrt 2 & -c/\sqrt 2 + s' s & -s - s' c/\sqrt 2 \\ 
s/\sqrt 2 & -s/\sqrt 2 - s' c & c - s' s/\sqrt 2 \end{array} \right] 
\left[ \begin{array} {c} \nu_1 \\ \nu_2 \\ \nu_3 \end{array} \right].
\end{equation}
Atmospheric $\nu_\mu \to \nu_\tau$ oscillations now occur with $\sin^2 2 
\theta = 4 s^2 c^2$ and $\Delta m^2 = 2 m_0 m_1 \simeq 4 \times 10^{-3}$ 
eV$^2$, and solar $\nu_e \to c \nu_\mu + s \nu_\tau$ vacuum oscillations 
occur with $\sin^2 2 \theta \simeq 1$ and $\Delta m^2 \simeq 4 \sqrt 2 s c 
s' m_0 I \simeq 4 \times 10^{-10}$ eV$^2$, if $m_0 = 2$ eV, $s = c = 
1/\sqrt 2$, and $s' = 0.01$.

\section{Conclusion}

The idea of nearly mass-degenerate neutrinos \cite{4,5} of a few eV may be 
cosmologically relevant as a component of dark matter \cite{6}.  If the 
origin of their common mass $m_0$ is the seesaw mechanism \cite{8}, then 
there is an irreducible splitting \cite{10} of the order $10^{-9}~m_0$ 
due to the charged-lepton masses.  This is very suitable for vacuum solar 
neutrino oscillations with $\Delta m^2 \sim 10^{-10}$ eV$^2$.  On the 
other hand, the above generic statement says nothing about atmospheric 
neutrino oscillations, but once the latter are incorporated (by hand or 
with the help of some new physics), then the residual splitting is 
available for solar neutrino oscillations.

\section*{Acknowledgments}
I thank the organizers Cesareo Dominguez and Raoul Viollier for their great 
hospitality at Cape Town.  This work was supported in part by the 
U.~S.~Department of Energy under Grant No.~DE-FG03-94ER40837.

\newpage
\begin{center}
\begin{picture}(339,100)(0,0)
\ArrowLine(21,0)(60,0)
\Text(40,-8)[c]{$\nu$}
\ArrowLine(60,0)(120,0)
\Text(90,-8)[c]{$l$}
\ArrowLine(120,0)(180,0)
\Text(150,-8)[c]{$\nu$}
\Text(180,0)[c]{$\times$}
\ArrowLine(240,0)(180,0)
\Text(180,-12)[c]{$N$}
\Text(210,-8)[c]{$\nu$}
\ArrowLine(300,0)(240,0)
\Text(270,-8)[c]{$l$}
\ArrowLine(339,0)(300,0)
\Text(320,-8)[c]{$\nu$}
\PhotonArc(150,-45)(101,26,154)58
\Text(150,68)[c]{$W$}
\PhotonArc(210,45)(101,206,334)58
\Text(210,-68)[c]{$W$}
\end{picture}
\vskip 1.2in
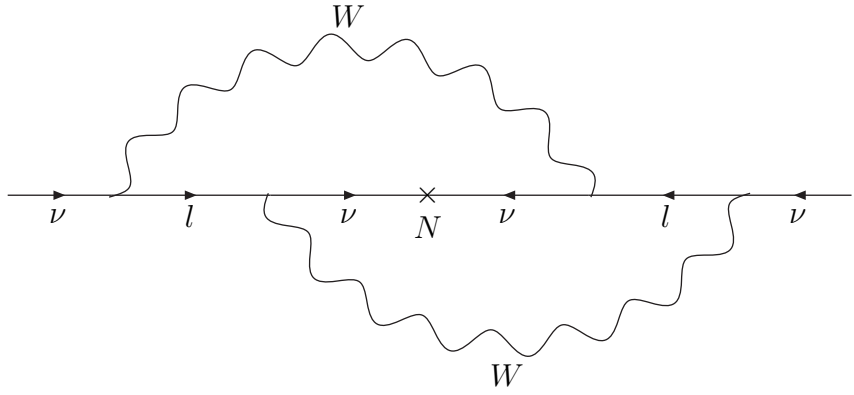
{\bf Fig.~1}. Two-loop radiative breaking of neutrino mass degeneracy. 
\end{center}

\end{document}